\documentclass[seceq,preprint]{ptptex}

\title{
Question on the Existence of Gravitational Anomalies
}

\author{
Mitsuo \textsc{Abe}$^{1,}$\footnote{E-mail: abe@kurims.kyoto-u.ac.jp}
and Noboru \textsc{Nakanishi}$^{2,}$\footnote{Professor Emeritus of 
Kyoto University. E-mail: nbr-nak@trio.plala.or.jp}%
}

\inst{
$^1$Research Institute for Mathematical Sciences, Kyoto University,\\ Kyoto 606-8502, Japan
\\
$^2$12-20 Asahigaoka-cho, Hirakata 573-0026, Japan
}

\notypesetlogo

\abst{
The existence of gravitational anomalies claimed by Alvarez-Gaum\'{e} and Witten 
is examined critically. It is pointed out that they were unaware of the essential 
difference between T-product quantities and T*-product quantities. Field equations 
and, therefore, the Noether theorem are, in general, violated in the case of
T*-product quantities, that is, those directly calculable from Feynman integrals. 
In the 2-dimensional case, it is explicitly confirmed that the 
energy-momentum tensor is strictly conserved if the above stated property of the
T*-product quantities is correctly taken into account. The non-existence of
gravitational anomalies is explicitly demonstrated for the BRS-formulated 
2-dimensional quantum gravity in the Heisenberg picture.
}

\begin{document}

\maketitle

\section{Introduction}
In 1984, Alvarez-Gaum\'{e} and Witten\cite{rf:1} (A.-G. and W.) claimed that there exist 
gravitational anomalies in the $(4k+2)$-dimensional spacetime ($k=0,1,2, \ldots$). 
In that work, they calculated the 1-loop Feynman integrals for the $(2k+2)$-point functions
of the energy-momentum tensor, and showed that the results are inconsistent
with the conservation law of the energy-momentum tensor.
The purpose of the present paper is to point out that their argument is not 
sufficient for establishing the existence of gravitational anomalies and that, in fact,
there exists no gravitational anomaly in the 2-dimensional 
case.\footnote{A preliminary report was made half a decade ago.\cite{rf:2}}

The flaw in their reasoning results from the fact
that they were careless with regard to the essential difference between the
T-product and the T*-product. The former is converted into the latter
when the Hamiltonian formalism is transcribed into the Lagrangian formalism.
The expressions appearing in the covariant perturbation theory and 
in the path-integral formalism are written as the vacuum expectation values of
the T*-product \textit{but not of the T-product.} 
Nevertheless, A.-G. and W. regarded the Feynman integrals 
as quantities written in terms of the T-product.  

The T-product is a sum over products of local operators multiplied 
by a product of $\theta$-functions of time differences; therefore,
in general, it becomes non-covariant when differentiated.
By contrast, the T*-product is defined in such a way that
time differentiations always act \textit{after} the vacuum
expectation value of the T-product of the canonical fields is taken.
Hence, T*-product quantities are always covariant. The price paid for
this bonus is that the T*-product is no longer a product in the mathematical
sense; a T*-product quantity involving a factor 0 is not necessarily 
equal to 0. Accordingly, Feynman integrals are not necessarily consistent
with field equations and, therefore, with the Noether theorem.
This violation of the Noether theorem should not be confused with
an anomaly, as it can be calculated explicitly from the difference between the 
T*-product and the T-product. 

As is well known, the Adler-Bell-Jackiw anomaly is established by calculating 
the Feynman integral corresponding to the triangle diagram.
The reason that the chiral anomaly is correctly obtained in this case is  that
the expression for the chiral current involves no differentiation.
Contrastingly, the expression for the energy-momentum tensor
necessarily contains time differentiations. 
In the T*-product quantity, those differentiations act from outside the
vacuum expectation value, but not directly on the fields in the expression for 
the energy-momentum tensor. This fact implies a
nontrivial difference between the T*-product and the T-product.
However no analysis of this difference is given in the work of A.-G. and W.

The main purpose of  the present paper is to show explicitly that,
at least in the 2-dimensional case, what A.-G. and W. called the gravitational
anomaly is nothing more than the contribution from the 
difference between the T*-product and the T-product. Thus, we establish that
the energy-momentum tensor is strictly conserved if correctly calculated.
We believe that this fact implies that in fact there exists no gravitational anomaly.

The present paper is organized as follows.
In \S2, we discuss the apparent violation of the Noether theorem encountered 
in the path-integral approach. 
In \S3, we consider the 2-dimensional Weyl field and show that
what A.-G. and W. interpreted as a gravitational anomaly 
is nothing more than the apparent violation of the conservation law
due to the use of the T*-product quantities.
In \S4, we examine the paper of  A.-G. and W. more closely,
because some people assert that even if the energy-momentum tensor
is conserved, the conclusion of A.-G. and W. would remain valid. 
With regard to this point, some comments
are made about the validity of the Virasoro anomaly.
Furthermore, we closely analyze the meaning of the general-coordinate
non-invariance that cannot be removed by local counter-terms.
In \S5, in order to demonstrate that there exists no gravitational anomaly, 
we consider BRS-formulated 2-dimensional quantum gravity 
coupled with the Weyl fields. The final section is devoted to discussion.
In the Appendix, some formulae for singular-function products are presented.

\section{T*-product and the Noether theorem}
In a previous paper, we investigated the pathological nature of the covariant 
perturbation theory and the path-integral formalism caused by 
the T*-product.\cite{rf:3} 
In this section, we first reproduce the general formula for the 
field-equation-violating contribution due to the T*-product
in the path integral.

The generating functional, $Z(J)$, of the Green functions
is formally expressed as a path integral:
\begin{equation}
Z(J)=\int \Big(\prod_A \mathcal{D} \varphi_A\Big)  
\exp i \int d^N x \Big(\mathcal{L} + \sum_A J_A \varphi_A \Big). 
\end{equation}
Here, $S=\int d^N x \mathcal{L}(x)$ is the action for the fields 
$\varphi_A (x)$ in the $N$-dimensional spacetime, 
$\mathcal{D} \varphi_A$ is the path-integral measure
normalized as $Z(0)=1$,  
and $J_A (x)$ denotes the source function for $\varphi_A (x)$.
Let $F(\varphi)$ be an arbitrary function of 
$\varphi_{A_1} (y_1), \ldots,$ $\varphi_{A_m} (y_m)$, and
let $\delta \varphi_A$ be a \textit{field-independent} variation of $\varphi_A$.
The path-integral measure should be invariant under the
functional translation $\varphi_A \to \varphi_A +\delta \varphi_A$.
Accordingly, by considering a variation of a particular field $\varphi_A$
in $F(i^{-1}\partial /\partial J)Z|_{J=0}$,
we obtain
\begin{equation}
i\Big\langle \mathrm{T}^* \frac{\delta}{\delta \varphi_A} S\cdot F(\varphi) 
\Big\rangle + \Big\langle \mathrm{T}^* \frac{\delta}
{\delta \varphi_A} F(\varphi) \Big\rangle = 0. 
\label{eq:2-2}
\end{equation}
This is the T*-product version of the field equation 
$(\delta /\delta \varphi_A) S =0$. The second term in (\ref{eq:2-2}) is 
the field-equation-violating term resulting from the use of the T*-product.
It is thus seen that we cannot naively use the field equations,
and therefore the Noether theorem, with the T*-product quantities.
Hence, we must treat the current conservation law very carefully
in the covariant perturbation theory.

Now, we consider the infinitesimal symmetry transformation 
\begin{equation}
\delta^{\epsilon} \varphi_A (x) \equiv \varphi'_A (x')
- \varphi_A (x).
\end{equation}
Thus, we define
\begin{equation}
\delta^{\epsilon}_* \varphi_A (x) \equiv \varphi'_A (x)
- \varphi_A (x), 
\end{equation}
so that 
\begin{equation}
\delta^{\epsilon} \varphi_A
= \delta^{\epsilon}_* \varphi_A + 
\delta^{\epsilon} x^{\mu} \cdot \partial_{\mu} \varphi_A, 
\end{equation}
where $ \delta^{\epsilon} x^{\mu} \equiv x'^{\mu} - x^{\mu} $. 

The Noether current $J^{\mu}$ is defined by
\begin{equation}
\epsilon J^{\mu} \equiv \sum_A \delta^{\epsilon}_* \varphi_A
\cdot \frac{\partial}{\partial(\partial_{\mu} \varphi_A)} \mathcal{L}
+ \delta^{\epsilon} x^{\mu} \cdot \mathcal{L}. 
\end{equation}
Then, the Noether identity is
\begin{equation}
\epsilon \partial_{\mu} J^{\mu} =
-\sum_A \delta^{\epsilon}_* \varphi_A \cdot 
\frac{\delta}{\delta \varphi_A}S + \delta^{\epsilon} \mathcal{L}, 
\end{equation}
where the last term on the right-hand side vanishes if the Lagrangian density is 
invariant under the symmetry transformation. However, the first term cannot be 
set to zero, because the field equations do not hold for the T*-product 
quantities, as stated above. In particular, for the energy-momentum
tensor $T^{\mu \nu}$, we have
\begin{equation}
\partial_{\mu} T^{\mu \nu} = - \sum_A \partial^{\nu} \varphi_A
\frac{\delta}{\delta \varphi_A}S. 
\end{equation}
Hence, with the help of (\ref{eq:2-2}), we obtain
\begin{equation}
\langle \mathrm{T}^* \partial_{\mu} T^{\mu \nu} (x) \cdot F(y) \rangle
= -i \Big\langle \mathrm{T}^* \sum_A \epsilon_{A}\, \frac{\delta}{\delta \varphi_A (x)}
[\partial^{\nu} \varphi_A (x) \cdot F(y)] \Big\rangle ,
\label{eq:2-9}
\end{equation}
where $\epsilon_{A}=1$ if $A$ is bosonic and $\epsilon_{A}=-1$ if $A$ is fermionic.

For illustration, we consider a free massive scalar field 
$\phi (x)$ in $N$ dimensions. Its field equation is
\begin{equation}
(\square + m^2)\phi = 0. 
\end{equation}
The energy-momentum tensor is given by
\begin{equation}
T^{\mu \nu} = \partial^{\mu} \phi \cdot  \partial^{\nu} \phi
- \frac{1}{2} \eta^{\mu \nu} (\partial^{\sigma} \phi \cdot
\partial_{\sigma} \phi - m^2 \phi^2), 
\end{equation}
which, of course, satisfies the conservation law
\begin{equation}
 \partial_{\mu} T^{\mu \nu} = (\square + m^2)\phi 
\cdot \partial^{\nu} \phi =0, 
\end{equation}
owing to the field equation. 
We should note, however, that the Feynman propagator, 
\begin{equation}
\langle \mathrm{T}^* \phi (x) \phi (y) \rangle  =
\langle \mathrm{T} \phi (x) \phi (y) \rangle  = \varDelta_F (x-y), 
\end{equation}
does not satisfy the Klein-Gordon equation but, instead, the equation
\begin{equation}
(\square +m^2) \varDelta_F (x-y) = -i \delta^N (x-y). 
\end{equation} 
Accordingly, by straightforward calculation, we find a \textit{nonvanishing} 
result for the divergence of the 2-point function of the 
energy-momentum tensor:
\begin{eqnarray}
\langle \mathrm{T}^* \partial_\mu T^{\mu \nu} (x) \cdot
T^{\lambda \rho} (y) \rangle &\equiv&
\partial^x_\mu \langle \mathrm{T}^* T^{\mu \nu} (x)
T^{\lambda \rho} (y) \rangle \nonumber\\  
&=&-i(\eta^{\lambda \sigma}  \partial^{\rho}
+\eta^{\rho \sigma}  \partial^{\lambda}
-\eta^{\lambda \rho}  \partial^{\sigma}) \partial^{\nu} \varDelta_F (x-y) \cdot
\partial_{\sigma} \delta^N (x-y) \nonumber\\
&&-im^2 \eta^{\lambda \rho} \partial^\nu 
\varDelta_F (x-y) \cdot \delta^N (x-y). 
\label{eq:2-15}
\end{eqnarray}
This is, of course, \textit{not} a gravitational anomaly.
Indeed, the right-hand side of (\ref{eq:2-9}) becomes
\begin{eqnarray}
 -i \Big\langle \mathrm{T}^* &&\frac{\delta}{\delta \phi(x)}
[\partial^\nu \phi(x) \cdot T^{\lambda \rho} (y)] \Big\rangle  \nonumber\\
=&&-i \langle \mathrm{T}^* \partial^\nu \phi(x) \,
[(-\partial^x_\sigma) \{ (\eta^{\lambda \sigma} \partial^\rho
+ \eta^{\rho \sigma} \partial^\lambda) \phi (y) \cdot \delta^N (x-y) \}  \nonumber\\
&&-\eta^{\lambda \rho} (-\partial^x_\sigma) \{ \partial^\sigma
\phi (y) \cdot \delta^N (x-y) \} + m^2 \eta^{\lambda \rho}
\phi(y) \delta^N (x-y)] \, \rangle, 
\end{eqnarray}
which is exactly equal to (\ref{eq:2-15}).

In order to avoid the obstruction caused by the T*-product, it is convenient to
define the anomaly in terms of Wightman functions.\cite{rf:4}
Let $j^{\mu}$ be a symmetry current;
then the anomaly for the corresponding symmetry exists if we have
\begin{equation}
\partial^x_{\mu}\langle j^{\mu}(x) \varphi_1 (y_1) 
\cdot \cdot \cdot \varphi_n (y_n) \rangle \neq 0 
\end{equation}
for some fields  $\varphi_1, \ldots , \varphi_n$ .
With regard to the gravitational anomaly of the above model, 
we have only to calculate the quantitly
\begin{equation}
\partial^x_{\mu} 
\langle 
T^{\mu\nu} (x) \phi (y_1) \phi (y_2) \rangle. 
\end{equation}
It is readily shown that this quantity vanishes using the Klein-Gordon equation 
for $\varDelta^{(+)} (x-y_j)$.
Thus, we conclude that the gravitational anomaly does not exist.

\section{Energy-momentum conservation in the Weyl theory}
We consider the 2-dimensional complex Weyl field $\psi (x)$. 
Its free-field action is given by
\begin{equation}
S=i \int d^2 x \psi^\dagger \partial_- \psi. 
\end{equation}
We employ the light-cone coordinates, $x^\pm = (x^0 \pm x^1)/\sqrt{2}$.
The field equation and the 2-dimensional anti-commutation relations are
\begin{eqnarray}
\partial_- \psi &=&0, \\
\{ \psi (x), \psi (y) \} &=& 0, \qquad \{ \psi (x), \psi^\dagger (y) \} 
=\delta (x^+ - y^+), 
\end{eqnarray}
respectively. Accordingly, the 2-point Wightman function and the
Feynman propagator are, respectively, as follows:
\begin{eqnarray}
\langle \psi (x) \psi^\dagger (y) \rangle &=&\frac{1}{2\pi i}
\cdot \frac{1}{x^+ -y^+ -i0},  \\
\langle \mathrm{T}^* \psi (x) \psi^\dagger (y) \rangle 
&=&\frac{1}{2\pi i} \cdot \frac{1}{x^+ -y^+ -i0(x^- - y^-)} \nonumber\\
&\equiv& \frac{1}{2\pi i} 
\left[ \frac{\theta (x^- - y^-)} {x^+ -y^+ -i0}
+ \frac{\theta (-x^- + y^-)} {x^+ - y^+ +i0} \right]. 
\end{eqnarray}
It is very important to write \textit{explicitly} the fact that 
the Feynman propagator depends on $x^- - y^-$.

The energy-momentum tensor $T^\mu_{\mspace{12mu} \nu}$ is given by
\begin{eqnarray}
T^-_{\mspace{12mu} +} = T_{++} &=&
\frac{i}{2}(\psi^\dagger \partial_+ \psi -
\partial_+ \psi^\dagger \cdot \psi), \label{eq:3-6} \\
T^+_{\mspace{12mu} +} = T_{-+} &=& 
\frac{i}{2}(- \psi^\dagger \partial_- \psi +
\partial_- \psi^\dagger \cdot \psi). \label{eq:3-7}
\end{eqnarray}
Although $T_{-+}$ vanishes owing to the field equation,
it \textit{cannot} be ignored in the calculation of the T*-product
quantities.

A straightforward calculation yields
\begin{eqnarray}
\langle \mathrm{T}^* T_{++} (x) T_{++} (y) \rangle &=&
\frac{1}{8\pi^2} \cdot \frac{1}{(x^+ -y^+ -i0(x^- - y^-))^4} \nonumber\\
&=&\frac{1}{8\pi^2} 
\left[ \frac{\theta (x^- - y^-)} {(x^+ -y^+ -i0)^4}
+ \frac{\theta (-x^- + y^-)} {(x^+ - y^+ +i0)^4} \right]. \label{eq:3-8}
\end{eqnarray}
Hence, we have
\begin{equation}
\partial^x_- \langle \mathrm{T}^* T_{++} (x) T_{++} (y) \rangle =
\frac{1}{24\pi i} \delta'''(x^+ - y^+) \delta(x^- - y^-). 
\label{eq:3-9}
\end{equation}

We can rewrite the above results into those in momentum space.
With the help of the formula
\begin{equation}
\int d^2 z \, \frac{\theta (\pm z^-)}{z^+ \mp i0}\, e^{ipz}
= -2\pi \cdot \frac{\theta (\mp p_+)}{p_- \mp i0},  
\end{equation}
we see that the Fourier transform of (\ref{eq:3-8}) is given by
\begin{equation}
\frac{i}{24\pi} \cdot \frac{p_+^{\mspace{12mu} 3}}{p_- + i0p_+}. 
\label{eq:3-11}
\end{equation}
It is important to write explicitly the infinitesimal imaginary part of the
denominator.
Differentiating (\ref{eq:3-8}) with respect to $x^-$ is equivalent to 
multiplying (\ref{eq:3-11}) by $-ip_-$, and therefore we obtain $(1/24 \pi)p_+^{\mspace{12mu} 3}$, 
which is, of course, the Fourier transform of (\ref{eq:3-9}). 
This quantity is identically that which A.-G. and W. regarded as the gravitational anomaly 
(see \S4).\cite{rf:1}
We can show, however, that it is merely the contribution from the 
difference between the T*-product quantity and the T-product quantity.

It is convenient to work in the spacetime representation. 
Dropping the terms proportional to $\epsilon(x^{-}-y^{-}) \delta(x^{-}-y^{-})=0$,
we obtain
\begin{eqnarray}
\langle \mathrm{T}^* &&T_{-+} (x) T_{++} (y) \rangle \nonumber\\
=&&\frac{1}{4\pi i} \left[ \mathrm{Pf} \frac{1}{(x^+ - y^+)^2}
\cdot \delta (x^+ - y^+) + \mathrm{Pf} \frac{1}{x^+ - y^+}
\cdot \delta' (x^+ - y^+) \right] \delta (x^- -y^-), \nonumber\\
\end{eqnarray}
where Pf denotes the ``finite part", i.e., $\mathrm{Pf} (1/z^n) \equiv \Re [1/(z-i0)^n]$. 
Hence, we find
\begin{eqnarray}
\partial^x_+ &&\langle \mathrm{T}^* T_{-+} (x) T_{++} (y) \rangle \nonumber\\
=&&\frac{1}{4\pi i} \left[ -2 \mathrm{Pf} \frac{1}{(x^+ - y^+)^3}
\cdot \delta (x^+ - y^+) + \mathrm{Pf} \frac{1}{x^+ - y^+}
\cdot \delta'' (x^+ - y^+) \right] \delta (x^- -y^-). \nonumber\\
\label{eq:3-13}
\end{eqnarray}
From (\ref{eq:3-9}) and (\ref{eq:3-13}), we obtain
\begin{eqnarray}
\langle \mathrm{T}^* &&[\partial_- T_{++} (x) +
\partial_+ T_{-+} (x)] T_{++} (y) \rangle \nonumber\\
=&& \frac{1}{4\pi i} \left[ -\frac{1}{6}
\delta'''(x^+ - y^+) - 4 \mathrm{Pf} \frac{1}{(x^+ - y^+)^3}
\cdot \delta (x^+ - y^+) \right] \delta (x^- - y^-), 
\label{eq:3-14}
\end{eqnarray}
where we have made use of the identity (\ref{eq:A-6}) presented in the Appendix.

Next, note that the quantity on the right-hand side of (\ref{eq:2-9}) becomes
\begin{eqnarray}
i\Big\langle \mathrm{T}^* &&\frac{\delta}{\delta \psi (x)}
[\partial_+ \psi (x) \cdot T_{++} (y)] \Big\rangle +
i \Big\langle \mathrm{T}^* \frac{\delta}{\delta \psi^\dagger (x)}
[\partial_+ \psi^\dagger (x) \cdot T_{++} (y)] \Big\rangle \nonumber\\
=&&\frac{1}{2\pi i} \left[ - \mathrm{Pf} \frac{1}{(x^+ - y^+)^2}
\cdot \delta' (x^+ - y^+) -2 \mathrm{Pf} \frac{1}{(x^+ - y^+)^3}
\cdot \delta (x^+ - y^+) \right] \delta (x^- -y^-). \nonumber\\
\label{eq:3-15}
\end{eqnarray}
With the help of the identity (\ref{eq:A-4}), we see that \textit{(\ref{eq:3-15}) 
coincides with (\ref{eq:3-14}).} Thus, we have shown that what 
A.-G. and W. called the gravitational anomaly
is simply the contribution from the T*-product.

Finally, we make a remark concerning the Majorana Weyl field $\psi$, used by
Green, Schwarz and Witten.\cite{rf:5} This field has only one field degree of freedom.
Hence, without introducing an extra field,
\textit{it cannot be quantized}, because there is
no canonical conjugate independent of $\psi$. Nevertheless, setting up the 
2-dimensional anticommutation relation
\begin{equation}
\{ \psi (x), \psi (y) \} =\delta (x^+ - y^+) 
\end{equation}
by hand and replacing (\ref{eq:3-6}) and (\ref{eq:3-7}) by the same expressions with 
the daggers removed, we can repeat the above analysis.
Doing so, we obtain the same results, except for the appearance of an overall
multiplicative factor of 2 \textit{on the left-hand side only}.
Thus, (\ref{eq:2-9}) does not hold in this model. The reason for
this is that the path integral does \textit{not} exist.\footnote{%
To write down the path integral, we expand $\psi$ as $\psi=\sum_i \theta_i
\varphi_i$, where the quantities $\theta_i$ are Grassmann numbers and we have $i\int d^2 x \,
\varphi_i \partial_- \varphi_j = \delta_{ij}$. Then the action vanishes because
$S=i \int d^2 x \, \psi \partial_- \psi$ $ =\sum_i \theta_i^{\; 2} =0$.
We cannot avoid this result by introducing conjugate Grassmann numbers 
$\theta_i^\dagger$, because $\psi$ satisfies a \textit{first-order} differential equation.
Indeed, if we do so, then the action becomes that of the \textit{complex}
Weyl field.
}

\section{Examination of the paper of A.-G. and W. }
In \S 3, we established that the energy-momentum tenser is strictly 
conserved, at least in the 2-dimensional case, if the difference between 
the T-product and the T*-product is correctly taken into account. 
However, some people believe that the claim of A.-G. and W. itself is valid nevertheless. 
To explicitly show that in fact it is not valid, 
we more closely examine the reasoning employed in the paper of A.-G. and W.
in this section. 

We first quote one paragraph, together with a footnote, of their paper.
\hfill\break

``With $\gamma_{-}\psi=\partial_{-}\psi=0$, the only non-vanishing component of 
the energy-momentum tensor is 
$T_{++}=\frac{1}{2}i\bar\psi\gamma_{+}\overleftrightarrow{\partial_{+}}\psi$,
and the linearized interaction of fermions with the gravitational field is $\Delta\mathcal{L}=
-h_{--}\frac{1}{4}i\bar\psi\gamma_{+}\overleftrightarrow{\partial_{+}}\psi$.
We will study the effective action to second order in the metric perturbation $h$, 
by studying the two-point function
\begin{equation*}
U(p) = \int d^{2}x\, e^{ip\cdot x} \langle \Omega \vert \mathrm{T}(T_{++}(x)T_{++}(0))
\vert \Omega \rangle.
\eqno(\textrm{A.-G. and W.}\ 11)
\end{equation*}
Now, it is possible to see without any computation that there must be an anomaly.
The naive conservation law for $T_{++}$ is $\partial_{-}T_{++}=0$; it leads to the naive
Ward identity $p_{-}U=0$.$^{\star}$
If true, this would imply $U=0$ for all non-zero $p_{-}$, and hence (by analyticity) for all $p_{-}$.
But $U$, as the two-point function of the hermitian operator $T_{++} $, cannot vanish.
So there must be an anomaly. \hfill\break
[footnote]\hfill\break
$^{\star}$ Naively there is no equal time commutator term in this Ward identity.
If one looks at (A.-G. and W. 11) as a two-point function in flat space, the anomaly we will find
can be regarded as an anomalous commutation relation 
$[T_{++}(x),\, T_{++}(y)]=(i/48\pi)\delta'''(x-y) +\,$tree
level terms. It is closely related to the anomaly in the Virasoro algebra in string theories.
But we will see that upon coupling $T_{++}$ to the gravitational field, the anomaly is 
a breakdown of general covariance.''
\hfill\break

In the succeeding paragraph, they calculate $U(p)$ using the Feynman
diagrammatic method and obtain
\begin{equation*}
U(p)=\frac{i}{24\pi} \frac{p_+^{\, 3}}{p_-}. 
\eqno(\textrm{A.-G. and W.}\ 13)
\end{equation*}
Then they proceed to discussing the ``question of the covariance 
of the effective action". As the effective action corresponding to
(A.-G. and W. 13), they consider 
\begin{equation*}
-\frac{1}{192\pi} \int d^2 p \frac{p_+^{\, 3}}{p_-}h_{--}(p)h_{--} (-p). 
\eqno(\textrm{A.-G. and W.}\ 15)
\end{equation*}
Under the general coordinate transformation of the external gravitational
field $h_{\mu \nu}(x)$, they show that (A.-G. and W. 15) cannot be made invariant
by adding a local functional of fields.

Now, in the paragraph quoted above, A.-G. and W. defined the 2-point function
$U(p)$ as a \textit{T-product} quantity. Nevertheless, in the succeeding 
paragraph, they calculated it by means of a \textit{T*-product} 
quantity. Evidently, they were unaware of the essential difference
between the T-product and the T*-product. Indeed, while they used 
the equation $T_{-+}=0$, which is \textit{not} valid in the case of the T*-product,
they calculated a quantity involving $T_{++}$ with the method for calculating
the T*-product quantity. Thus their reasoning is clearly internally inconsistent.

It is quite interesting to analyze the naive argument appearing below (A.-G. and W. 11),
in which A.-G. and W. indicated the existence of an anomaly: Why did they reach 
an invalid conclusion in spite of the fact that they considered T-product
quantities in this paragraph?\footnote{One of the authors (N. N.)
would like to thank Prof. T. Kugo for raising this question.}
The reason can be clarified by examining the footnote presented there.
For this purpose, we reproduce the description concerning the
Virasoro anomaly in a book of Green, Schwarz and 
Witten.\cite{rf:5} It is in essence as follows:\footnote{In their notation, 
spacetime coordinates are denoted by $(\tau, \sigma)$ and 
$\sigma^\pm =\tau \pm \sigma$, without the factor of $1/\sqrt{2}$.}
\hfill\break

``We consider a 2-dimensional massless scalar field $\phi$, which
satisfies
\begin{equation}
\partial_+ \partial_- \phi =0. 
\end{equation}
The ++ component of its energy-momentum tensor is
\begin{equation}
T_{++} = \partial_+ \phi \cdot \partial_+ \phi, 
\end{equation}
which satisfies the conservation law
\begin{equation}
\partial_- T_{++} =0. 
\end{equation}
But, because the T-product is noncommutative with the time differentiation,
we obtain
\begin{equation}
\partial_- \langle \mathrm{T} \, T_{++} (\sigma, \tau)
 T_{++} (\sigma', \tau') \rangle
=\frac{1}{2} \delta (\tau -\tau') \langle
[ T_{++} (\sigma, \tau), \, T_{++} (\sigma', \tau') ] \rangle. 
\label{eq:4-4}
\end{equation}
Perturbative calculation yields
\begin{equation}
\langle \mathrm{T} \, T_{++} (\sigma, \tau) T_{++} (\sigma', \tau') \rangle
= \frac{1}{8} \cdot \frac{1}{(\sigma^+ - \sigma'^+)^4}. 
\label{eq:4-5}
\end{equation}
Substituting (\ref{eq:4-5}) into (\ref{eq:4-4}) and making use of the formula
\begin{equation}
\partial_- \frac{1}{\sigma^+} =i\pi \delta (\sigma) \delta (\tau), 
\label{eq:4-6}
\end{equation}
we find 
\begin{equation}
\langle [ T_{++} (\sigma, \tau), \, T_{++} (\sigma', \tau) ] \rangle
=-\frac{i\pi}{24} \delta'''(\sigma -\sigma'), 
\label{eq:4-7}
\end{equation}
which is the equal-time \textit{anomalous} commutator."
\hfill\break

In the above, it is (\ref{eq:4-5}) that is mistaken: It is a perturbative result, and
therefore it is a T*-product quantity. 
Because differentiation commutes with the T*-product,
it cannot be substituted into the formula (\ref{eq:4-4}), which is a formula for
the T-product. The curious formula (\ref{eq:4-6}) is correctly written
\begin{equation}
\partial_- \frac{1}{\sigma^+ -i0 \sigma^-}
 =i\pi \delta (\sigma) \delta (\tau). 
 \end{equation}
Correspondingly, the correct version of (\ref{eq:4-5}) is
\begin{equation}
\langle \mathrm{T}^* T_{++} (\sigma, \tau) T_{++} (\sigma', \tau') \rangle
= \frac{1}{8} \cdot \frac{1}{[\sigma^+ - \sigma'^+ -
i0(\sigma^- - \sigma'^-)]^4}. 
\end{equation}

It is easy to correctly carry out the calculation if one employs the
Wightman functions. 
Noting the relations
\begin{eqnarray}
\langle T_{++} (\sigma, \tau) T_{++} (\sigma', \tau') \rangle
&=& \frac{1}{8} \cdot \frac{1}{(\sigma^+ - \sigma'^+ -i0)^4}, \\
\langle T_{++} (\sigma', \tau') T_{++} (\sigma, \tau) \rangle
&=& \frac{1}{8} \cdot \frac{1}{(\sigma^+ - \sigma'^+ +i0)^4}, 
\end{eqnarray}
we have
\begin{eqnarray}
\langle [ T_{++} (\sigma, \tau), \,  T_{++} (\sigma', \tau')] \rangle
&=& \frac{1}{8} \left[ \frac{1}{(\sigma^+ - \sigma'^+ -i0)^4}
-\frac{1}{(\sigma^+ -\sigma'^+ +i0)^4} \right] \nonumber\\
&=&-\frac{i\pi}{24} \delta'''(\sigma^+ - \sigma'^+). \label{eq:4-12}
\end{eqnarray}
In particular, when $\tau=\tau'$, (\ref{eq:4-12}) reduces to (\ref{eq:4-7}), but
(\ref{eq:4-12}) is \textit{not} anomalous. Note that $\delta'''$ can be
expressed in terms of $\delta'$, as is seen in (\ref{eq:A-4}).

Thus, we see that A.-G. and W. omitted the commutator term
because they misunderstood it to be anomalous.

It is also noteworthy that A.-G. and W. regarded the violation of the conservation 
law of $T_{\mu \nu}$ as evidence of a gravitational anomaly.
Their consideration of the effective action amounts to no more than simply checking 
whether or not  (A.-G. and W. 15) can be made invariant by adding some local terms. 
However, such an investigation is completely unnecessary, because the quantity in 
(A.-G. and W. 15) itself does not appear.

There is the assertion\footnote{The authors would like to thank Prof.\ T. Kugo 
for detailed discussions about this standpoint.} 
that the ``gravitational anomaly'' should be defined \textit{not} in terms of the inevitable 
violation of translational invariance \textit{but} in terms of the existence of 
generally-noncovariant terms in the path integral generally-covariantized
by introducing an external gravitational field $g_{\mu\nu}(x)$.  
The (logarithm of the) path integral $S_{\mathrm{eff}}$ is expanded into powers of 
$h_{\mu\nu}(x)=g_{\mu\nu}(x)-\eta_{\mu\nu}$, and one considers its variation 
(nontrivial lowest order in $h_{\mu\nu}$ only) under an infinitesimal general coordinate 
transformation. That is, what one considers is the infinitesimal variation of 
\begin{equation}
S_{\mathrm{eff}}^{\;(2)}= \mathrm{const}\int d^N x \int d^N y \; h_{\mu \nu} (x) \langle \mathrm{T}^*
T^{\mu \nu} (x) T^{\lambda \rho} (y) \rangle h_{\lambda \rho} (y) \label{eq:4-13}
\end{equation}
under the transformation $\delta_{*}^{\;\epsilon} h_{\mu \nu} (x)=
-\partial_{\mu}\epsilon_{\nu}(x)-\partial_{\nu}\epsilon_{\mu}(x)$.
Integrating by parts, we encounter the quantity
$\langle \mathrm{T}^{*}\partial_{\mu}T^{\mu\nu}(x)\cdot T^{\lambda\rho}(y)\rangle$.

The above formulation does not straightforwardly apply to the case of the 
Weyl field, because the action for the Weyl field cannot be generally covariantized 
by using $g_{\mu\nu}$, as it is necessary to introduce the \textit{zweibein}.
Although A.-G. and W. defined the zweibein as $e_{\mu a} = \eta_{\mu a} + (1/2)h_{\mu a}$, 
this form represents nothing but the unwarranted omission of the
antisymmetric part of $e_{\mu a}$. Indeed, the transformation implied by this
definition explicitly violates the general-coordinate invariance of the action.
The zweibein transforms not as a tensor but as two vectors under 
general-coordinate transformations. Therefore, the expansion must be carried out in powers
not of $h_{\mu\nu}$ but of $\tilde h_{\mu a }=e_{\mu a}-\eta_{\mu a}$.

The starting action must be invariant under general-coordinate transformations.
The invariant action for the Weyl field is given by
\begin{equation}
S_{W}=\int d^{2}x (e_{+-}j_{-} - e_{--}j_{+}), 
\label{eq:4-14}
\end{equation}
with $j_{\pm}=\pm T_{\pm +}$, where $T_{\pm +}$ is given by (\ref{eq:3-6})
and (\ref{eq:3-7}). We can explicitly confirm its invariance under the infinitesimal
general-coordinate transformation defined by 
$\delta_{*}^{\;\epsilon} e_{\pm -}=-\partial_{\pm}\epsilon^{\lambda}\cdot e_{\lambda -}
-\epsilon^{\lambda}\partial_{\lambda}e_{\pm -}$ and $\delta_{*}^{\;\epsilon}\psi
=-\epsilon^{\lambda}\partial_{\lambda}\psi$. [Note that the integrand of (\ref{eq:4-14})
can be written $\epsilon^{\mu\nu}e_{\mu -}j_{\nu}$.]

Thus, for the Weyl field, (\ref{eq:4-13}) should be replaced by
\begin{equation}
S_{\mathrm{eff}}^{\;(2)}= \mathrm{const}\int d^2 x \int d^2 y \; \tilde h_{\mu -} (x) \langle \mathrm{T}^*
T^{\mu}{}_{+} (x) T^{\lambda}{}_{+} (y) \rangle \tilde h_{\lambda -} (y), \label{eq:4-15}
\end{equation}
with $\delta_{*}^{\;\epsilon} \tilde h_{\mu a}=-\partial_{\mu}\epsilon_{a}(x)$. 
Only the term for $\mu=\lambda=-$ is nontrivial; in momentum space, it
yields (A.-G. W. and 15). The other three terms are local terms; 
they would vanish if the T*-product were not taken. In momentum space, they are
quadratically divergent. After regularization, they can be written
\begin{equation}
\mathrm{const}\int d^{2}p [ -p_{+}^{2} \tilde h_{--}(p)\tilde h_{+-}(-p)
- p_{+}^{2} \tilde h_{+-}(p)\tilde h_{--}(-p) + p_{+}p_{-}\tilde h_{+-}(p)\tilde h_{+-}(-p)].
\label{eq:4-16}
\end{equation}
By choosing the coefficient appropriately, we see that the sum of (A.-G. and W. 15) and 
(\ref{eq:4-16}) is invariant under the transformation
$\delta_{*}^{\;\epsilon}\tilde h_{\mp -}(p)= -ip_{\mp} \epsilon_{-}(p)$. 

Up to this point, everything is essentially the same as in the 2-dimensional massless
scalar field theory, in which $\langle \mathrm{T}^{*}T^{\mu\nu}(x) T^{\lambda\rho}(y)\rangle$
has a nonlocal term proportional to 
\begin{equation}
\frac{p^{\mu}p^{\nu}p^{\lambda}p^{\rho}}{p^{2}+i0}
\end{equation}
in momentum space. Its divergence is, of course, a local term. 
This fact is well known as the ``conformal anomaly''.\cite{rf:6}

The qualitative difference between the presently considered case and the scalar-field case 
arises when one requires local Lorentz invariance for the Weyl theory. 
While $S_{W}$ in (\ref{eq:4-14}) is invariant,
$S_{\mathrm{eff}}^{\;(2)}$ in (\ref{eq:4-15}) is \textit{not}.
Indeed, the infinitesimal local Lorentz transformation of $e_{\mu -}$ yields
$\delta \tilde h_{--}=0$ and $\delta \tilde h_{+-}=\tilde\epsilon$ 
at lowest order. Hence (A.-G. and W. 15) is already invariant at lowest order,
but any local term depending on $\tilde h_{+-}$ is \textit{not}.
However, we must note that this non-invariance is caused by the fact that the free 
action of the Weyl field is not invariant under the local Lorentz transformation. 
In general, under a symmetry transformation that does not leave the free action
invariant, the invariance is not preserved at each order in the perturbation theory.

Shortly after A.-G. and W., Langouche\cite{rf:7} and Leutwyler\cite{rf:8} resolved
this problem by exactly carrying out the path integral of the effective action.
According to Leutwyler, $S_{\mathrm{eff}}$ is consistent with the
general-coordinate invariance but not with the local-Lorentz invariance, and the 
local-Lorentz non-invariance cannot be transferred into the general-coordinate
non-invariance in an admissible way. Thus, although a Lorentz anomaly may exist, 
it cannot be claimed that a ``gravitational anomaly'' exists in the exact expression for 
$S_{\mathrm{eff}}$.

According to Leutwyler, the energy-momentum tensor $T^{\mu\nu}$, defined in 
a frame-independent manner, does not satisfy the \textit{general-covariant}
conservation law; $\nabla_{\mu}T^{\mu\nu}$ is a local polynomial, but it cannot be
removed by adding a local polynomial to $T^{\mu\nu}$.
However, the general-covariant conservation law $\nabla_{\mu}T^{\mu\nu}=0$
is important only in classical gravity; it has no physical significance in quantum
gravity. Indeed, we emphasize that \textit{the existence of a ``gravitational
anomaly'' in this sense has nothing to do with the obstruction to the unitarity
of the physical S-matrix of quantum gravity.}

\section{2-dimensional gravitational theory}
A precise treatment of the gravitational anomaly must be made in the framework
of quantum gravity. We emphasize that when the gravitational field is quantized, 
there is no general-coordinate invariance.  
\textit{The gravitational anomaly is nothing but an anomaly with respect to
translational invariance}\footnote{
Here, of course, the ``translation" is defined as
a (pseudo-)unitary transformation
of quantum fields, under which the effective action is \textit{always} non-invariant.
One might say that the effective action is translationally invariant under
a linear approximation with respect to $h_{\mu \nu}$, but this merely implies
that it is ``translationally invariant"
\textit{if any arbitrary function can be regarded as a constant}.
Genuine translational invariance must be checked for the \textit{exact}
solution in the framework of quantum gravity.
} \textit{in the framework of quantum gravity.}

In this section, in order to explicitly demonstrate that there is no gravitational anomaly, 
we consider the BRS-formulated 
conformal-gauge 2-dimensional quantum gravity coupled with $D$ Weyl fields.
If it is coupled with $D$ scalar fields instead, the model can be interpreted as
a string theory in $D$-dimensional spacetime.
Previously, we thoroughly investigated this model and found the complete
solution in terms of Wightman functions.\cite{rf:9,rf:10,rf:11,rf:12} \ Extension to the case 
of Weyl fields is straightforward.

In the conformal gauge, the gravitational field $g^{\mu \nu}$ is 
parametrized as $g^{\pm \mp}=\exp (-\theta)$ and $g^{\pm \pm} =
\exp (-\theta) h_\pm$. Hence, to first order, the zweibein $e_{\mu a}$
is given by $e_{\pm \mp}=\exp (\frac{1}{2} \theta)$ and
$e_{\pm \pm}= -\frac{1}{2} \exp (\frac{1}{2} \theta) h_\mp$
in the symmetric gauge $e_{\mu a}=e_{a \mu}$.

The action for $D$ Weyl fields coupled to the zweibein is given 
by\footnote{For $D=1$, (\ref{eq:5-1}) reduces to (\ref{eq:4-14}).}
\begin{equation}
S_W = \frac{i}{2} \int d^2 x \, \left[\tilde{\psi}^\dagger_M
\left(e_{+-} \partial_- \tilde{\psi}_M
- e_{--} \partial_+ \tilde{\psi}_M\right)
-\left(e_{+-} \partial_- \tilde{\psi}^\dagger_M
- e_{--} \partial_+ \tilde{\psi}^\dagger_M\right) \cdot \tilde{\psi}_M \right], 
\label{eq:5-1}
\end{equation}
where the sum over $M=1, \ldots, D$ is understood.
Setting  $\tilde{\psi}_M = \exp (-\frac{1}{4} \theta) \psi_M$,
we have
\begin{equation}
S_W = \frac{i}{2} \int d^2 x \, \left[\psi^\dagger_M
\left(\partial_- \psi_M + \frac{1}{2} h_+ \partial_+ \psi_M\right)
-\left(\partial_- \psi^\dagger_M + \frac{1}{2} h_+ 
\partial_+ \psi^\dagger_M\right) \cdot \psi_M \right] 
\label{eq:5-2}
\end{equation}
to first order.

Let $\tilde{b}^\mu$, $c^\mu$ and $\overline{c}^\mu$ be
the B field, the FP ghost and the FP anti-ghost, respectively.
The BRS transformation, which is, of course, nilpotent, is defined by
\begin{eqnarray}
\delta_{*} h_{\pm} 
&=& 2\partial_{\mp}c^{\pm} +(\partial_{\pm}c^{\pm}-\partial_{\mp}c^{\mp})\cdot h_{\pm}
-c^{\lambda}\partial_{\lambda} h_{\pm} - \frac{1}{2}\partial_{\pm}c^{\mp}\cdot h_{\pm}^{2}, 
\label{eq:5-3}\\
\delta_{*}\psi_{M} 
&=& -\frac{1}{2}\partial_{+}c^{+}\cdot \psi_{M} - c^{\lambda}\partial_{\lambda}\psi_{M}
+\frac{1}{4}\partial_{+}c^{-}\cdot h_{+}\psi_{M}, \\
\delta_{*}c^{\mu}
&=& -c^{\lambda}\partial_{\lambda}c^{\mu}, \qquad
\delta_{*}\bar c^{\mu} = i\tilde b^{\mu}, \qquad
\delta_{*}\tilde b^{\mu} = 0.
\end{eqnarray}
The Weyl action (\ref{eq:5-2}) is exactly BRS invariant, in spite of the fact that we have
adopted a non-covariant gauge. Because the BRS transformation (\ref{eq:5-3}) of $h_{\pm}$
is the same as that given previously, except for the second-order terms, 
the action proper to the zweibein is essentially the same as in our previous 
work.\cite{rf:9}
Then, from (\ref{eq:5-2}), we see that the total action is given by
\begin{equation}
S = \int d^2 x \, (\mathcal{L}_0 + \mathcal{L}_I),
\end{equation}
with
\begin{eqnarray}
\mathcal{L}_0 &=& -\frac{1}{2} \tilde{b}^+ h_+ 
- i \overline{c}^+ \partial_- c^+ + (+ \leftrightarrow -) 
+ \frac{i}{2} \left[ \psi^\dagger_M \partial_- \psi_M 
-\partial_- \psi^\dagger_M  \cdot \psi_M \right], \\
\mathcal{L}_I &=& \frac{1}{2} h_+ [ -2i \overline{c}^+ 
\partial_+ c^+ - i(\partial_+ \overline{c}^+ \cdot c^+
+ \partial_- \overline{c}^+ \cdot c^-)] + (+ \leftrightarrow -)
+\frac{1}{2} h_+ T_{++} +O(h^2), \nonumber\\
\label{eq:5-8}
\end{eqnarray}
where
\begin{equation}
T_{++} \equiv \frac{i}{2} (\psi^\dagger_M \partial_+ \psi_M
- \partial_+ \psi^\dagger_M \cdot \psi_M)  
\end{equation}
and $(+ \leftrightarrow -)$ indicates to interchange  \textit{attached letters}
in the preceding expression.

In the operator formalism, higher-order terms, i.e., $O(h^2)$,
yield no contribution, because of the field 
equation $h_\pm =0$, and moreover the left-moving mode and
the right-moving mode decouple completely, because of the field equations
$\partial_\mp c^\pm =0$, $\partial_\mp \overline{c}^\pm =0$, etc.
It should be noted that such simplicity is \textit{never} realized in the
path-integral formalism, because the T*-product violates the field equation.
The terms of first order in $h_\pm$ yield the B-field equations:
\begin{eqnarray}
\mathcal{T}^+ &\equiv&  -\tilde{b}^+  - 2i\overline{c}^+ \partial_+ c^+ 
- i\partial_+ \overline{c}^+ \cdot c^+ + T_{++} =0, \\
\mathcal{T}^- &\equiv& -\tilde{b}^- - 2i\overline{c}^- \partial_- c^- 
- i\partial_- \overline{c}^- \cdot c^- =0.
\end{eqnarray}

From this point, everything is carried out as in our previous work,
except for the part in which the Weyl fields are relevant.
For this reason, we do not describe our reasoning and proceed directly 
to the Wightman functions involving the Weyl fields.
The nonvanishing $n$-point truncated Wightman functions involving the 
Weyl fields are only those which consist of one $\psi_M$, one $\psi^\dagger_M$
and $(n-2)$ $\tilde{b}^+$ fields ($n \geqq 2$). 
Their explicit expressions are 
\begin{eqnarray}
\langle \psi_M&&(x_1) \tilde{b}^+ (x_2) \cdot \cdot \cdot \tilde{b}^+ (x_{n-1})
\psi^\dagger_N (x_n) \rangle_\mathrm{T} \nonumber\\
=&& \delta_{MN} \left( \frac{i}{2} \right)^{n-2}  \frac{1}{(2\pi i)^{n-1}}
\sum^{(n-2)!}_{P(j_2, \ldots, j_{n-1})}
(\partial^R_{j_2} - \partial^L_{j_2})(\partial^R_{j_3} - \partial^L_{j_3})
\cdots(\partial^R_{j_{n-1}} - \partial^L_{j_{n-1}}) \nonumber\\
&& \cdot \left( \frac{1}{x_1^+ - x_{j_2}^+ -i0} 
\cdot \frac{1}{x_{j_2}^+ - x_{j_3}^+ \mp i0} 
\cdot \cdot \cdot \frac{1}{x_{j_{n-2}}^+ - x_{j_{n-1}}^+ \mp i0}
\cdot \frac{1}{x_{j_{n-1}}^+ - x_{n}^+ -i0} \right), \nonumber\\
\end{eqnarray}
with similar expressions for other orderings of field operators.
Here, $P(j_2, \ldots, j_{n-1})$ denotes a permutation of ($2, \ldots,n-1$),
we have $x_j^+ - x_k^+ \mp i0 = x_j^+ - x_k^+ - i0$ for $j<k$ and 
$= x_j^+ - x_k^+ +i0$ for $j>k$, and $\partial^{R/L}_j$ acts only on the
$x_j^+$ appearing in the right/left factor.\footnote{For example, 
$\partial^R_j [f(x_j^+)g(x_j^+)] \equiv f(x_j^+)\partial_j g(x_j^+)$.}

\textit{All truncated Wightman functions are consistent with translational invariance.} 
The fact that there is no gravitational anomaly
is easily seen as follows. The Noether currents for translational invariance are
\begin{eqnarray}
J^-_{\mspace{12mu} +} &=& -i\overline{c}^+ \partial_+ c^+ + T_{++}, \\
J^+_{\mspace{12mu} -} &=& -i\overline{c}^- \partial_- c^-, 
\end{eqnarray}
with $J^\pm_{\mspace{12mu} \pm} = 0$. Because any nonvanishing truncated
vacuum expectation value of a product of $J^-_{\mspace{12mu} \nu} (x)$ 
and fundamental fields
$\varphi_1 (y_1), \ldots, \varphi_n (y_n)$ depends only on 
$x^+, y_1^{\mspace{12mu} +}, \ldots, y_n^{\mspace{12mu} +}$,
but not on $x^-$, we trivially have
\begin{equation}
\partial_-^x 
\langle J^-_{\mspace{12mu} \nu} (x)
\varphi_1 (y_1) \cdot \cdot \cdot \varphi_n (y_n) \rangle = 0. 
\end{equation}
We have a similar expression for $J^+_{\mspace{12mu} \nu}$. 
Thus, there is no gravitational anomaly (see \S2).
 
However, this model is not free of all anomalies.
As in the scalar-field case,\cite{rf:9} \ 
the B-field equations exhibit a ``field-equation anomaly", namely,
a slight violation of field equations at the representation level.
Explicitly, we find that
\begin{eqnarray}
\langle \tilde{b}^+ (x_1) \mathcal{T}^+ (x_2) \rangle 
&=& -\langle \mathcal{T}^+ (x_1) \mathcal{T}^+ (x_2) \rangle 
= (D-26)\varPhi^{++} (x_1^+ - x_2^+), \label{eq:5-16}\\
\langle \tilde{b}^- (x_1) \mathcal{T}^- (x_2) \rangle 
&=& -\langle \mathcal{T}^- (x_1) \mathcal{T}^- (x_2) \rangle 
= -26 \, \varPhi^{++} (x_1^+ - x_2^+), \label{eq:5-17}
\end{eqnarray}
where
\begin{equation}
\varPhi^{++} (z^+) \equiv \frac{1}{8\pi^2} \cdot
\frac{1}{(z^+ - i0)^4}, 
\end{equation}
just as in the scalar-field case. The perturbation-theoretical
counterparts of (\ref{eq:5-16}) and (\ref{eq:5-17}) are the unexpected 1-loop 
contributions to $\langle \mathrm{T}^* \tilde{b}^\pm (x_1) \tilde{b}^\pm (x_2) \rangle $,
which are easily calculated by using (\ref{eq:5-8}) according to the Feynman rules; 
the reason for their presence is that
the Feynman propagator $\langle \mathrm{T}^*
\tilde{b}^\pm (x) h_\pm (y) \rangle$ is \textit{nonvanishing},
owing to the fact that the T*-product violates the field equation.

The Noether currents for translational invariance
 contain an anomalous contribution from the
above field-equation anomaly at the representation level, but
we can define the anomaly-free currents as
\begin{equation}
\tilde{J}^\mp_{\mspace{12mu} \pm} \equiv J^\mp_{\mspace{12mu} \pm} - \mathcal{T}^\pm
- i\partial_\pm (\overline{c}^\pm c^\pm) = \tilde{b}^\pm,
\end{equation}
so that the anomaly-free translational generators are given by
\begin{equation}
P_\pm \equiv \int dx^\pm \tilde{b}^\pm. 
\end{equation}
We can define the anomaly-free BRS current similarly.\cite{rf:9}

\section{Discussion}
In the present paper, we have shown that the quantities calculated by A.-G. and W.
do not directly show the existence of gravitational 
anomalies. In order to establish the existence of gravitational anomalies,
it is necessary to calculate T-product quantities, \textit{not 
T*-product quantities directly calculable using Feynman integrals.}  
We have explicitly demonstrated that,  in the 2-dimensional case,
the energy-momentum tensor is strictly conserved 
if the fact that the T*-product quantities violate the field equation 
is correctly taken into account.
Although it is very difficult to demonstrate this explicitly
in higher-dimensional cases, it is clear that the reasoning of A.-G. and W.
is insufficient for establishing the existence of gravitational anomalies.
Note that A.-G. and W. also presented a naive argument from which they inferred 
the existence of gravitational anomalies in higher-dimensional cases from their
result in the 2-dimensional case. It is therefore natural to conjecture 
that the \textit{non-existence} of a gravitational anomaly in higher-dimensional cases 
follows from \textit{that} in the 2-dimensional case. 
It is interesting to note that the cancellation
of the 10-dimensional gravitational anomaly is known as the anomaly-free
condition of the superstring theory proposed by Green and Schwarz.

We emphasize the fundamental importance of strictly distinguishing 
T*-product quantities from T-product quantities. 
In recent studies, most calculations are carried out using T*-product quantities,
but it seems that they are done without being aware of the fact that
field equations and the Noether theorem do not hold in the case of  
T*-product quantities. It is generally quite dangerous to investigate
symmetry properties on the basis of Feynman diagrammatic calculations.
As it is much more difficult to calculate T-product quantities
than T*-product quantities, because the former are, in general,
non-covariant, it is preferable to investigate symmetry properties
by means of Wightman functions in the Heisenberg picture.

\appendix 
\section*{Singular Function Identities} 
In this appendix, we present some identities for products of singular functions.  

Taking the imaginary part of the self-evident identity
\begin{equation}
\left[ \frac{1}{(z-i0)^{n+1}} \right]^2 =
\frac{1}{(z-i0)^{2n+2}}, 
\end{equation}
we obtain
\begin{equation}
\mathrm{Pf} \frac{1}{z^{n+1}} \cdot \delta^{(n)} (z)
= \frac{(-1)^{n+1} n!}{2 \cdot (2n+1)!} \delta^{(2n+1)} (z). 
\end{equation}
In particular, for $n=0$ and for $n=1$, this becomes
\begin{equation}
\mathrm{Pf} \frac{1}{z} \cdot \delta (z)
= - \frac{1}{2} \delta' (z), 
\label{eq:A-3}
\end{equation}
and
\begin{equation}
\mathrm{Pf} \frac{1}{z^2} \cdot \delta' (z)
= \frac{1}{12} \delta''' (z), 
\label{eq:A-4}
\end{equation}
respectively. 

Differentiating (\ref{eq:A-3}) twice, we obtain
\begin{equation}
2 \mathrm{Pf} \frac{1}{z^3} \cdot \delta (z)
-2 \mathrm{Pf} \frac{1}{z^2} \cdot \delta' (z)
+\mathrm{Pf} \frac{1}{z} \cdot \delta'' (z)
= -\frac{1}{2} \delta''' (z). 
\label{eq:A-5}
\end{equation}
Adding (\ref{eq:A-4}) twice to (\ref{eq:A-5}), we have
\begin{equation}
\mathrm{Pf} \frac{1}{z} \cdot \delta'' (z)
= -\frac{1}{3} \delta''' (z) - 2 \mathrm{Pf} \frac{1}{z^3} \cdot \delta (z). 
\label{eq:A-6}
\end{equation}

%


\begin{thebibliography}{99}
\bibitem{rf:1}
L. Alvarez-Gaum\'{e} and E. Witten, \NPB{234,1984,269}.
\bibitem{rf:2}
N. Nakanishi, Soryushiron Kenkyu (Kyoto) \textbf{100} (1999), 167.
\bibitem{rf:3}
M. Abe and N. Nakanishi, \PTP{102,1999,1187}.
\bibitem{rf:4}
M. Abe and N. Nakanishi, 
Int. J. Mod. Phys. A \textbf{11} (1996), 2623.
\bibitem{rf:5}
M. B. Green, J. H. Schwarz and E. Witten, \textit{Superstring Theory}, 
Vol. I (Cambridge University Press, 1987), Chap. 3.
\bibitem{rf:6}
D. W. D\"usedau, 
Phys. Lett. B \textbf{188} (1987), 51.
\bibitem{rf:7}
F. Langouche, 
Phys. Lett. B \textbf{148} (1984), 93.
\bibitem{rf:8}
H. Leutwyler, 
Phys. Lett. B \textbf{153} (1985), 65.
\bibitem{rf:9}
M. Abe and N. Nakanishi, 
Int. J. Mod. Phys. A \textbf{14} (1999), 521.
\bibitem{rf:10}
M. Abe and N. Nakanishi, 
Int. J. Mod. Phys. A \textbf{14} (1999), 1357.
\bibitem{rf:11}
N. Nakanishi, in \textit{Proceedings of the 30th International
Conference on High Energy Physics, Osaka, 2000} Vol. II,
ed. C. S. Lim and T. Yamanaka (World Scientific, 2001), p.1402. 
\bibitem{rf:12}
N. Nakanishi, \PTP{111,2004,301}.
\end{thebibliography}
\end{document}